\documentclass[%
 aip,
 pop,%
 amsmath,amssymb,
 reprint,%
]{revtex4-2}

\usepackage{graphicx}
\usepackage{dcolumn}
\usepackage{bm}
\usepackage[colorlinks=true,allcolors=blue]{hyperref}
\usepackage{xcolor}
\usepackage{cancel} 
\usepackage{relsize} 

\newcommand*{\rom}[1]{\expandafter\@slowromancap\romannumeral #1@}

\begin{document}



\title{Uncertainty analysis of the plasma impedance probe}

\author{J. W. Brooks}
\email{john.brooks@nrl.navy.mil}
\affiliation{ 
U.S. Naval Research Laboratory, Washington, D.C., USA}

\author{M. C. Paliwoda}
\affiliation{ 
	U.S. Naval Research Laboratory, Washington, D.C., USA, Karle Fellow}


\date{\today}

\begin{abstract}
	
	A plasma impedance probe (PIP) is a type of in-situ, radio-frequency (RF) probe that is traditionally used to measure plasma properties (e.g. density) in low-density environments such as the Earth's ionosphere.  We believe that PIPs are underrepresented in laboratory settings, in part because PIP operation and analysis has not been optimized for signal-to-noise ratio (SNR), reducing the probe's accuracy, upper density limit, and acquisition rate.   
	This work presents our efforts in streamlining and simplifying the PIP design, model, calibration, and analysis for unmagnetized laboratory plasmas, in both continuous and pulsed PIP operation.  The focus of this work is a Monte Carlo uncertainty analysis, which identifies operational and analysis procedures that improve SNR by multiple orders of magnitude.  Additionally, this analysis provides evidence that the sheath resonance (and not the plasma frequency as previously believed) sets the PIP's upper density limit, which likely provides an additional method for extending the PIP's density limit.  
	

\end{abstract}


\maketitle

\section{Introduction}

	The Plasma Impedance Probe (PIP)\cite{harp1964,rafalskyi2015, kim2016, blackwell2005rsi} is an in-situ plasma diagnostic that is underutilized when compared with the Langmuir Probe (LP)\cite{mottsmith1926, merlino2007}. 
	Both probes share much in common: both are metal electrodes, often have similar geometries (e.g. planar and spherical), and form sheaths when placed in a plasma.  
	When the sheath is large with respect to the probe size, both probes struggle with accuracy~\cite{brooks2023,lobbia2017}.
	When electrically biased, both provide a measurement of the coupled sheath-plasma impedance, $Z$, around the probes, and fitting sheath-plasma models to these measurements provide many of the same plasma properties. These properties include: the electron plasma frequency,
	\begin{equation} \label{eq:wp}
	\begin{split}
		\omega_p =   \sqrt{\frac{n e^2}{m_e \epsilon_0}},
	\end{split}
	\end{equation}
	electron density ($n$)\cite{blackwell2005rsi},
	temperature ($T$)\cite{georgin2024,walker2008}, Debye length ($\lambda_D$), sheath thickness ($t_{sh}$)\cite{brooks2023,kim2014}, and plasma potential ($V_p$)\cite{blackwell2005pop}.  Historically, PIPs are  primarily used to measure density, but have also measured electron damping ($\nu$).  However, isolating the species-dependent collisional and collisionless terms within $\nu$ is still an active research topic in the PIP community\cite{walker2006, you2016, oberrath2018}.  
	
	Arguably the largest differences between PIPs and LPs are in how they are electrically operated and what they subsequently measure.  
	LPs are biased across a range of large excitation voltages ($V_{exc} > k_b T/e$) and swept at low rates compared with the ion and electron plasma frequencies.  They therefore only measure electrical resistance, Re($Z$).  Due to the large voltage sweeps, LP measurements encounter three distinct sheath regimes: unsaturated, ion-saturated, and electron-saturated.  The complexity of the underlying models and subsequent challenges in numerical analysis often lead to total probe errors between 10\% to 50\% \cite{lobbia2017}.  
	In contrast, PIPs~\cite{blackwell2005rsi, brooks2023} are excited across a range of frequencies ($\omega_{exc}$) above the ion-plasma frequency and near the electron-plasma frequency ($\omega_{p}$); therefore PIPs measure the complex impedance spectra, $Z(\omega)$, and ignore ion contributions.  Additionally, PIPs are excited at relatively lower voltages 
	which avoids saturated sheaths and allows for linearization of the Boltzmann relation.  We argue that these factors lead to a simpler PIP model which results in improved measurement accuracy over the LP.  
	
	Physically, PIPs operate by measuring the frequency response of the plasma~\cite{blackwell2005rsi, brooks2023}.  When placed in plasma, a sheath forms between the PIP and the bulk plasma.  Both the sheath and plasma regions have their own frequency-dependent dielectrics, $\epsilon_{sh}(\omega)$ and $\epsilon_{p}(\omega)$, respectively, which are dependent on their respective plasma properties.  These dielectrics electrically couple with the PIP and modify the PIP's electrical impedance, $Z_{PIP}(\omega)$.  An example of this coupling is it introduces two resonances ($\omega_{\pm}$) into the PIP's impedance.  
	Measuring $Z_{PIP}(\omega)$ allows us to infer the two dielectrics and their plasma properties.  Experimentalists typically perform this measurement by transmitting one of two types of waveforms to the PIP: swept waveforms and pulsed waveforms.  
	Swept waveforms~\cite{blackwell2005rsi,hopkins2014,you2016,oberrath2018}, the traditional method, use a vector network analyzer (VNA) to transmit continuous frequency sweeps and measure both the outbound and reflected waves.  This approach is generally recommended over the pulsed approach because it is simpler to setup/operate and provides a higher signal-to-noise ratio (SNR).  However, it has a slower acquisition rate, typically no faster than $100$~Hz.  
	The pulsed method~\cite{spencer2019,dubois2021,brooks2023}, a more modern approach, uses a signal generator to transmit a series of single, broad-spectrum pulses separated by a finite time delay and an oscilloscope to measure the voltages and currents associated with each pulse. This approach allows for $>1$ MHz acquisition rates but at the cost of lower SNR and higher experimental complexity~\cite{brooks2023}.  Although SNR has greater impact on the pulsed waveform method, it is a limiting factor for both methods.
	
 	We attribute the PIP's under-representation in the laboratory to several factors~\cite{rafalskyi2015,brooks2023}: PIP's higher cost, complexity, relatively low upper-density limit ($<$$10^{16}$~m$^{-3}$), and the long heritage of LPs.  
	This upper-density limit is set by several factors, including the underlying assumptions of the PIP's model~\cite{brooks2023}, but it is often attributed to the maximum resolvable frequency in a $Z_{PIP}(\omega)$ measurement.  Historically, the community has believed that PIP measurements must resolve frequencies up to the plasma frequency for unmagnetized plasmas and up to the upper hybrid frequency for magnetized plasmas.  Measuring higher densities, which scale with $\omega_p^2$ (Eq.~\ref{eq:wp}), becomes increasingly complicated as RF instrumentation cost, complexity, and sensitivity to calibration errors all increase appreciably with frequency.  This, in part, is why PIPs have been historically developed for lower-density environments, specifically the ionosphere where $n < 10^{12}$ m$^{-3}$.  
	Examples include sounding rockets\cite{jackson1959,steigies2000,spencer2008,suzuki2010,barjatya2013,patra2013,spencer2019}, 	satellites\cite{oya1979, oya1986}, and on the International Space Station \cite{wright2008,barjatya2009, amatucci2019}.  To a lesser extent, PIPs have also been used in lower-density laboratory plasmas, including DC discharges~\cite{sen1973,brooks2023}, Hall thruster plumes~\cite{bilen1999,hopkins2014}, and plasma processing applications~\cite{kokura1999} where $n < 10^{16}$ m$^{-3}$.   
		
	The goal of this work is to increase SNR in order to: i) improve overall measurement quality, ii) extend the upper density limit, iii) increase acquisition rates, and iv) allow the unit to be more accessible in the laboratory environment.
	To achieve this, we present here our work in modernizing and streamlining the PIP's model, design, operation, and analysis. 
	In Section~\ref{sec:model}, we discuss our updated PIP monopole design and accompanying analytical model.  
	In Section~\ref{sec:methods}, we discuss our streamlined calibration, operation, and analysis for both the continuous and pulsed PIP methods.  
	In Section~\ref{sec:uncertainty}, we perform a Monte Carlo uncertainty analysis that both quantifies the improvement in SNR from the previous two sections and provides recommendations for optimized PIP operation and analysis.

\section{Design and models  \label{sec:model}}

	To extract meaningful results from a measurement of $Z_{PIP}$, we require a physical model.  Below, we A) introduce our PIP-monopole design which allows for easier modeling and calibration, B) derive the accompanying PIP models for unmagnetized plasmas, and C) discuss their assumptions and limitations.  	
	
	\subsection{Design \label{subsec:pip_design}}
\begin{figure}
			\includegraphics[]{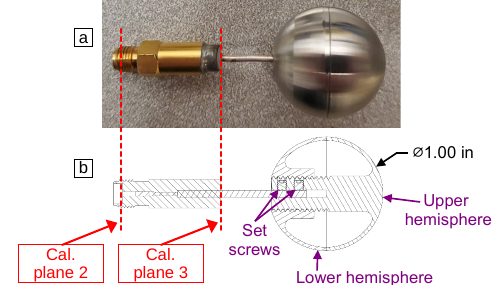}
			\caption{\label{fig:PIP_monopole} a) PIP-monopole.  b) CAD drawing.  Calibration planes 2 and 3 are discussed in Sec.~\ref{subsec:calibration}.  }
		\end{figure}

		For unmagnetized plasmas, we recommend the spherical monopole design as it is non-directional (unlike a dipole~\cite{balmain1964}) and is simpler to both construct and model.  Figure~\ref{fig:PIP_monopole} shows a recent monopole antenna design.		
		This probe was fabricated from a Pasternack RG401 semi-rigid coax cable where the outer conductor and dielectric were trimmed back to expose the inner conductor.  
		The head was a two-piece, machined SS316 hollow sphere, and the inner conductor was attached with set screws.  The head for our previous design~\cite{brooks2023} was a drilled aluminum sphere that was press fit onto the inner conductor.  
		
	\subsection{Models \label{subsec:PIP_monopole_model}}
	
		The model presented below is a reformulation of Blackwell's model~\cite{blackwell2005rsi} and closely follows our previous derivation~\cite{brooks2023}.
		
		Similar to previous work~\cite{buckley1966, balmain1966, blackwell2005rsi, blackwell2015,brooks2023}, we begin our model by using a lumped-element circuit framework and modeling the PIP's head and surrounding environment as one or more capacitors in series.  To clarify, we ignore inductance within the PIP's head but not in the surrounding environment.  The impedance spectra of any capacitor is $Z(\omega) = 1 / j \omega C$, and because of the spherical geometries, we use spherical-shell capacitor models, which have capacitance $C=4 \pi \epsilon (1/r_1 - 1/r_2)^{-1}$ with $r_1$ and $r_2$ being the radii of the inner and outer  electrodes, respectively.  
						
\begin{figure}
			\includegraphics[]{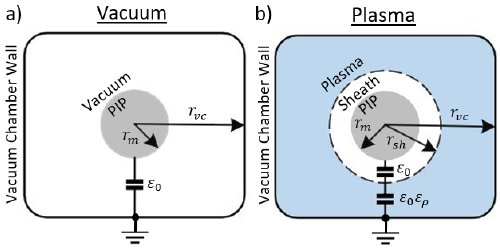}
			\caption{\label{fig:pip_model} Two circuit models of the PIP-monopole.  a) The vacuum model uses a single vacuum-filled capacitor.  b) The plasma model uses two capacitors in series: a vacuum dielectric for the sheath and a plasma dielectric for the plasma ($\epsilon = \epsilon_0 \epsilon_p$).}
		\end{figure}

		When no plasma is present (Figure~\ref{fig:pip_model}a), we model the vacuum around the PIP as a single capacitor, where the inner electrode is the PIP-monopole's head with radius $r_m$, the outer electrode is the grounded vacuum chamber wall with approximate radius $r_{vc}$, and dielectric $\epsilon=\epsilon_0$.  Recognizing that $r_m / r_{vc}  \approx 0$, the PIP's vacuum impedance is therefore 			
				
		\begin{subequations}
			\begin{align}
				Z_{vac} = & \frac{1}{4\pi j \omega \epsilon_0 r_m} .
				\intertext{We further simplify this expression to}
				\label{eq:Z_vacuum}
				 Z'_{vac} & =  \frac{ 1}{j \omega'}
			\end{align}
		\end{subequations}
		by introducing a normalized impedance, $Z' \equiv Z/Z_m$, a characteristic impedance of the monopole, 
		\begin{equation} \label{eq:Z_m}
			\begin{split}
				Z_{m} (r_m, \omega_p) \equiv \frac{1}{4\pi \epsilon_0 r_m \omega_p},
			\end{split}
		\end{equation}
		and a normalized frequency,  $\omega' \equiv \omega / \omega_p$.  
		
		When plasma is present (Figure~\ref{fig:pip_model}b), we model the sheath and plasma regions as two concentric capacitors in series.  The inner conductor is the monopole's head, the outer conductor is the vacuum vessel, and the middle conductor is the sheath-plasma boundary with radius, $r_{sh}$.  
		
		Because the sheath's density is much less than the bulk plasma, we approximate its dielectric as a homogeneous vacuum (e.g. $\epsilon \approx \epsilon_0$) with a normalized sheath thickness,
			\begin{equation} \label{eq:t_sh_prime}
			\begin{split}
				t'_{sh} \equiv \frac{r_{sh} - r_m}{r_{sh}} = \frac{t_{sh}}{r_{sh}}.
			\end{split}
		\end{equation}
		The sheath's impedance is therefore
		\begin{equation} \label{eq:Z_sheath}
			\begin{split}
				Z'_{sh} &= \frac{1}{j\omega'} t'_{sh}
			\end{split}
		\end{equation}
		where we have similarly normalized the impedance.

		For the bulk plasma, we model its dielectric as a plasma that is homogeneous, cold, collisional, and unmagnetized.  Specifically, $\epsilon = \epsilon_0 \epsilon_p$, where
		\begin{equation} \label{eq:epilson_p}
			\begin{split}
				\epsilon_{p} &=  \left(1 - \frac{1 }{\omega' \left( \omega' - j \nu' \right) } \right),
			\end{split}
		\end{equation}
		and $\nu' \equiv \nu/\omega_p$ is a normalized electron damping rate.  The inclusion of $j \nu' \omega'$ effectively introduces a damped, inductive term into this dielectric~\cite{blackwell2005rsi}.  The bulk plasma's impedance is
				
		\begin{equation} \label{eq:Z_plasma}
			\begin{split}
				Z'_{pl} &= \frac{1}{j\omega' } \frac{(1 - t'_{sh})}{\epsilon_p}.
			\end{split}
		\end{equation}
		
		The total PIP impedance is the sum of the sheath and bulk plasma impedances, 
		\begin{equation} \label{eq:Z_pip}
			\begin{split}
				Z'_{pip}  & = Z'_{sh} + Z'_{pl} \\
				& =  \frac{1}{j \omega'} \left( t'_{sh} +   \frac{\left(1 - t_{sh}'\right)}{\epsilon_p} \right) \\
				& = \frac{\nu'\omega' (1 - t'_{sh}) -   j (\omega'^2 - \omega'^2_{+})(\omega'^2 - \omega'^2_{-})}{\omega' \left( \nu'^2 \omega'^2 + (\omega'^2 - 1)^2 \right)}.
			\end{split}
		\end{equation}
		In electrical systems, resonances occur where Im($Z$)=0, and Eq.~\ref{eq:Z_pip} has two positive resonances~\cite{blackwell2005rsi,brooks2023}, 
		\begin{equation} \label{eq:Z_mono_zeros}
			\begin{split}
				\omega'_{\pm}
				&= \sqrt{ \frac{1 + t'_{sh} -\nu'^2 \pm \sqrt{ \left(1 + t'_{sh} - \nu'^2\right)^2 - 4 t'_{sh}} }{2} }.
			\end{split}
		\end{equation}
		We refer to the lower resonance  ($\omega_-$) as the sheath resonance because it disappears as the sheath vanishes,
		\begin{equation} \label{eq:sheath_resonance}
			\begin{split}
				\lim_{t'_{sh} \rightarrow 0} \omega_{-} = 0,
			\end{split}
		\end{equation}
		and we refer to $\omega_+$ as the damped-plasma resonance because it converges to $\omega_p$ as damping vanishes,
		\begin{equation} \label{eq:plasma_resonance}
			\begin{split}
				\lim_{ \nu' \rightarrow 0} \omega_{+} = \omega_p.
			\end{split}
		\end{equation}
		Further solving Eq.~\ref{eq:Z_mono_zeros}, we find that both resonances merge and disappear with large damping~\cite{gillman2018,brooks2023}, 	
		\begin{equation} \label{eq:Z_zeros_condition}
			\begin{split}
				\nu' \geq 1 - \sqrt{t'_{sh}}.
			\end{split}
		\end{equation}

		We also create a second PIP model by subtracting the vacuum impedance from Eq.~\ref{eq:Z_pip}, 
		\begin{equation} \label{eq:Z_diff}
			\begin{split}
			Z'_{diff} &= Z'_{pip} - Z'_{vac}\\
			& = \frac{1}{j \omega'}   \left( \frac{1}{\epsilon_p }- 1\right) \left(1 - t_{sh}' \right) \\ %
			& = \frac{\nu'\omega' (1 - t'_{sh}) -   j ( 1 - t'_{sh} ) (\omega'^2 - 1)}{\omega' \left( \nu'^2 \omega'^2 + (\omega'^2 - 1)^2 \right)}.
			\end{split}
		\end{equation} 
		Both Eqs.~\ref{eq:Z_pip} and~\ref{eq:Z_diff} have the same real component.  However, Eq.~\ref{eq:Z_diff} has a resonance at
		\begin{equation} \label{eq:diff_resonance}
			\begin{split}
				\omega_{diff}=\omega_p,
			\end{split}
		\end{equation}
		which isolates $\omega_p$ from $\nu'$ and $t'_{sh}$.
		
\begin{figure}
			\includegraphics[]{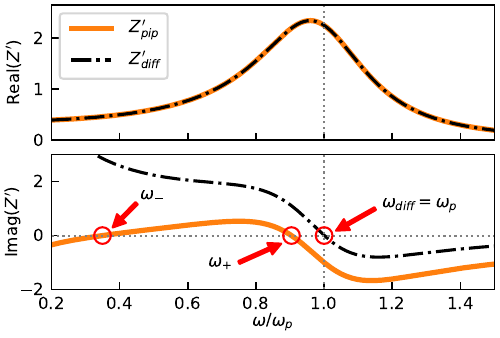}
			\caption{\label{fig:Z_prime_models} The real and imaginary components of both models ($Z'_{pip}$ and $Z'_{diff}$) are shown for $t_{sh}'=0.1$ and $\nu'=0.4$.  The three resonances are indicated.  }
		\end{figure}

		To better understand $Z'_{pip}$ and $Z'_{diff}$, we plot both in Figure~\ref{fig:Z_prime_models} for $t_{sh}'=0.1$ and $\nu'=0.4$.  We choose these values because i) they are values we have encountered experimentally and ii) the three resonances ($\omega_{\pm}$ and $\omega_{diff}$) are distinctly visible in the imaginary components.
		The real components have a peak near the plasma frequency and a width that broadens with higher damping~\cite{hopkins2014}.  

		
		Above, we presented two models of the PIP when plasma is present ($Z_{pip}$ and $Z_{diff}$), and we recommend using $Z_{pip}$ for most cases (discussed further in Sec.~\ref{sec:uncertainty}).  However, there are two cases where $Z_{diff}$ may be preferred.  
		First, $Z_{diff}$ is less dependent on $\nu'$ and is therefore its measurements are less susceptible to noise when damping is high ($\nu'$ $\scriptstyle \gtrsim$ 1).  
		This dependency on $\nu'$ is apparent when comparing Eq.~\ref{eq:Z_pip} and Eq.~\ref{eq:Z_diff}, while considering Eq.~\ref{eq:Z_mono_zeros}. The resonant frequencies ($\omega_{\pm}$ ) are heavily dependent upon $\nu'$, which are present in $Z_{pip}$ but not in $Z_{diff}$. At high damping, $\omega_+$ and $\omega_-$ merge and disappear per Eq.~\ref{eq:Z_zeros_condition}, while $\omega_p$ is isolated from $\nu'$ and $t'_{sh}$ per Eq.~\ref{eq:diff_resonance}. 
		Second, we resolved $Z_{diff}$ using planar and  cylindrical capacitors (in addition to spherical), and all three results provide the same $\omega_{diff}=\omega_p$ resonance.  This suggests that this result may be independent of probe geometry (i.e. by subtracting the vacuum measurement, we may be partially calibrating out the probe's geometry.)  This implies that in cases where the probe's geometry is difficult to model, it may be possible to identify $\omega_p$ by locating  Im$(Z_{diff})=0$ in a measurement alone (i.e. without the need of a model).  
		
		Despite building our models ($Z_{vac}$, $Z_{pip}$, and $Z_{diff}$) in electrical impedance ($Z$), we instead prefer doing analysis with the reflection coefficient~\cite{caspers2011,pozar2011},
		\begin{equation} \label{eq:Z_to_Gamma}
			\begin{split}
				\Gamma(\omega) = \frac{Z(\omega) - Z_0}{Z(\omega) + Z_0}.
			\end{split}
		\end{equation}
		In this expression, $Z_0$ is the characteristic impedance of the transmission lines, typically 50 $\Omega$. We convert $Z$ to $\Gamma$ using Eq.~\ref{eq:Z_to_Gamma}
		for both PIP measurements (vacuum and plasma) and our three PIP models, which results in $\Gamma_{vac}$, $\Gamma_{pip}$, and $\Gamma_{diff}$.
		We prefer $\Gamma$ because it results in lower uncertainty (see Sec.~\ref{sec:uncertainty}).  
		Converting from $Z$ to $\Gamma$ does not change the location of the three resonances: $\omega_{\pm}$ and $\omega_{diff}$.

	\subsection{PIP limitations and errors}


\section{Methods \label{sec:methods}}

\begin{figure}
		\includegraphics[]{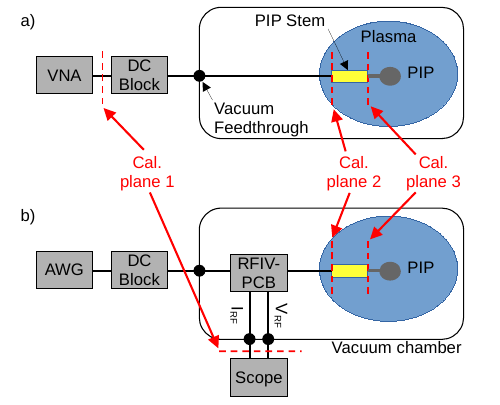}
		\caption{\label{fig:swept_and_pulsed_setups} Diagram of the PIP setup for: a) Swept PIP operation using a VNA, and b) pulsed PIP operation using an AWG and oscilloscope. }
	\end{figure}

	With the models established in Sec.~\ref{sec:model}, we next turn to PIP setup, calibration, and analysis for both swept and pulsed methods.  Both setups, diagrammed in Figure~\ref{fig:swept_and_pulsed_setups}, have much in common despite apparent differences, and we discuss both below.  	
	
	\subsection{Swept setup}
		The most notable feature of the swept PIP setup (Figure~\ref{fig:swept_and_pulsed_setups}a) is that it uses a VNA as both the stimulus and the measuring device, which simplifies PIP operation.   Typical VNA acquisition rates are relatively slow and are determined by several factors, including hardware limits and the amount of averaging required to achieve a reasonable SNR.  As a rough approximation, a higher-end VNA will have a maximum sweep rate of order 100~Hz assuming no averaging and measuring $10^3$ samples~\cite{coppermountain2018}.  
		A VNA operates by transmitting low-voltage frequency sweeps to the PIP  through a series of transmission lines (TLs).  
		Upon arriving at the PIP, the voltage is attenuated, partially reflected, and phase shifted due, in part, to the dielectric properties of the sheath and plasma regions.  The VNA measures both the reflected ($V_{refl}$) to transmitted ($V_{trans}$) voltages, and the ratio is the definition of the reflection coefficient~\cite{caspers2011,pozar2011},
		
		\begin{equation} \label{eq:Gamma_definition}
			\begin{split}
				\Gamma(\omega) \equiv \frac{V_{refl}(\omega)}{V_{trans}(\omega)}.
			\end{split}
		\end{equation}
		
		In this setup, the TLs include a DC-blocking filter which protects the VNA from DC voltages and allows the PIP's head to electrically float.  
		The coaxial cabling  is RG316 with SMA connectors, securely routed, and kept at a constant temperature.  Small changes in cable routing and temperature can result in calibration drift and therefore measurement error, especially at higher frequencies~\cite{czuba2011}.

	\subsection{Swept calibration \label{subsec:calibration}}
		The TL's between the VNA and PIP add unwanted attenuation and phase shift to the PIP measurement.  Calibration is the process of removing the contribution of the TLs and isolating the contribution of the PIP and surrounding dielectrics.  In the RF engineering community, this is referred to as ``moving the calibration plane''. 
		PIP calibration consists of two steps and therefore has three calibration planes (cal. planes) shown in Figures~\ref{fig:PIP_monopole} and~\ref{fig:swept_and_pulsed_setups}.  
		
		To assist with calibration, we intentionally built our PIP-monopole (Sec.~\ref{subsec:pip_design})  from a commercial, semi-rigid, coaxial cable, which provides two benefits.  i) It provides a convenient connector (e.g. SMA) that makes calibrating from cal. plane~1 to cal. plane~2 relatively simple.  ii) The stem of the PIP (the intact portion of the semi-rigid cable) can be modeled as a lossless transmission line, using the cable's published dielectric properties for calibrating from cal. plane~2 to cal. plane~3.  
 
		The first step (cal. plane 1 to 2) calibrates the transmission lines between the VNA and PIP's SMA connector.  In this work, we use a 1-port error model~\cite{janjusevic_2019, brooks2023} which requires measurements of 3 RF standards at both cal. plane 1 (we refer to these as the ``truth'' measurements) and at cal. plane 2.  Modern VNAs have built-in one-port error calibration procedures, but for our work, we perform the measurements manually and use \emph{scikit-rf}'s \emph{OnePort} calibration function\cite{scikitrf2023}.
		
		The second step (cal. plane 2 to 3) calibrates the PIP's stem.  Because there is no convenient connector (e.g. SMA) at cal. plane 3 to connect RF standards, we instead model the PIP's stem using a two-port, lossless transmission line model~\cite{caspers2011, brooks2023} and the published dielectric constant of the stem.  For our work, we build the model in \emph{scikit-rf} and invert the model to de-embed (calibrate) the stem as described in App.~\ref{app:2port_cal}.  
		
		App.~\ref{app:calibration_example} shows an example calibration using both steps.
		
		

	\subsection{Swept analysis}
	
		After acquiring and calibrating a $\Gamma_{pip}$ measurement, the next step is to fit the model to the measurement in order to solve for the model's three unknowns: $\omega_p$, $\nu$, and $t_{sh}$.  This requires several steps.
		First, we normalize $\omega_p$ in Eq.~\ref{eq:Z_pip} so that its value is roughly the same order as $\nu'$ and $t'_{sh}$, and therefore the fitting routine weights each parameter more equally.  
		Second, we convert our $Z_{pip}$ model (Eq.~\ref{eq:Z_pip}) into $\Gamma_{pip}$ by applying Eq.~\ref{eq:Z_to_Gamma} to the model.  
		Third, we trim our calibrated measurement to a finite frequency range, which is discussed in Sec.~\ref{sec:uncertainty}.  
		Fourth, we fit our $\Gamma_{pip}$ model to our $\Gamma_{pip}$ measurement using a minimization function (specifically, \emph{scipy}'s \emph{minimize} or \emph{least\_squares} functions) with a residual function that includes both the real and imaginary components of $\Gamma$.  We use the same procedure above when fitting with $Z_{diff}$.

\begin{figure}
			\includegraphics[]{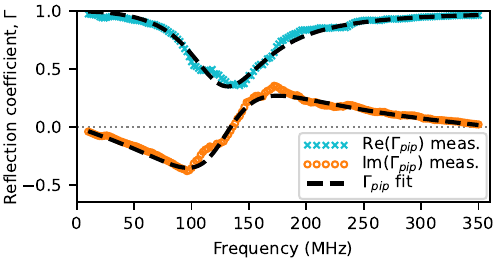}
			\caption{\label{fig:example_fit}  An example swept measurement is fit with the $\Gamma_{pip}$ model.  The calculated plasma parameters are $\omega_p / 2 \pi= 373$~MHz, $n = 1.73\cdot 10^{15}$ m$^{-3}$, $\nu'=0.151$ ($\nu = 353\cdot 10^6$ $s^{-1}$), and $t'_{sh} = 0.126$ ($t_{sh} = 0.91$~mm).  }
		\end{figure}

		Figure~\ref{fig:example_fit} shows an example of the $\Gamma_{pip}$ model fit to a calibrated, swept $\Gamma_{pip}$ measurement.    We attribute discrepancies between the model and the fit to the simplicity of model, which includes: ignoring plasma gradients, ignoring nearby electrical conductors (i.e. the PIP's stem), and using a simplistic sheath model.
		
	\subsection{Pulsed setup \label{subsec:pulsed_setup}}
	
		The pulsed PIP setup, shown in Figure.~\ref{fig:swept_and_pulsed_setups}b, uses an AWG (arbitrary waveform generator) to transmit sequential, preprogrammed waveform pulses to the PIP.  A custom circuit board (RFIV board)~\cite{brooks2023} isolates signals proportional to the pulsed voltage ($V_{RF}$) and currents ($I_{RF}$), which are measured by an oscilloscope in the time domain.  Because we are using a single, broad-spectrum pulse to measure the PIP's impedance (instead of a frequency sweep), we are able to achieve significantly higher acquisition rates ($>1$ MHz) but at the cost of lower SNR.  
		
\begin{figure}
			\includegraphics[]{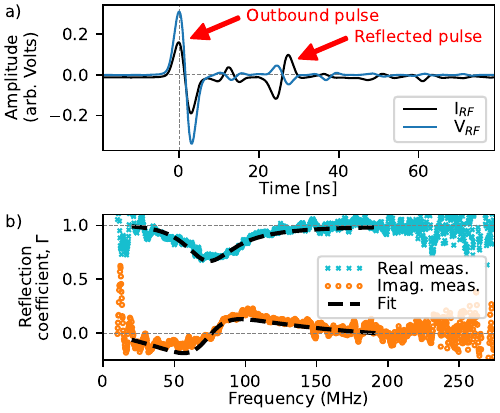}
			\caption{\label{fig:RFIV_pulses_realdata} An example pulsed measurement and fit using the $\Gamma_{pip}$ model.  a) Shows the raw oscilloscope measurements of $I_{RF}$ and $V_{RF}$.  The outbound pulses occur at 0~ns, and the reflected pulses, due to line length, occur 25~ns later. b) Shows the processed and calibrated $\Gamma_{pip}$ measurement and fit.  Fit results: $\omega_p / 2\pi=194$ MHz, $n=4.67\cdot10^{14}$ m$^{-3}$, $\nu'=0.185$ ($\nu=225 \cdot 10^{6}$~s$^{-1}$), and $t'_{sh}=0.148$ ($t_{sh}=1.60$~mm).}
		\end{figure}

		For our work, the AWG transmits a Gaussian monopulse,
		\begin{equation} \label{eq:monopuse}
			\begin{split}
				%
				g_{mp}(t; a, \sigma_{mp}) = \frac{a t}{\sigma_{mp}^2} \text{exp}\left(-\frac{1}{2}\left(\frac{t}{\sigma_{mp}}\right)^2\right),
			\end{split}
		\end{equation}
		which is the time derivative of a Gaussian distribution.  Here,  $a$ and $\sigma_{mp}$ are the  amplitude and width of the Gaussian distribution, respectively.   Figure~\ref{fig:RFIV_pulses_realdata}a shows an example measurement of both measured outbound and reflected monopulses for both $V_{RF}$ and $I_{RF}$.   
		The measured impedance of a single pulse at the scope is 
		\begin{equation} \label{eq:IV_to_Z}
			\begin{split}
				Z(\omega)=\frac{\text{FFT} \left\{ V_{RF}(t) \right\} } {\text{FFT}\left\{I_{RF}(t)\right\} }
			\end{split}
		\end{equation}
	 	where FFT$\left\{\, \right\}$ is the fast Fourier transform.  $Z(\omega)$ is then converted to $\Gamma(\omega)$ with Eq.~\ref{eq:Z_to_Gamma}.  Figure~\ref{fig:RFIV_pulses_realdata}b shows the calibrated $\Gamma_{pip}$ measurement and subsequent fit (discussed in Sec.~\ref{subsec:pulsed_analysis}).
				 			
\begin{figure}
			\includegraphics[]{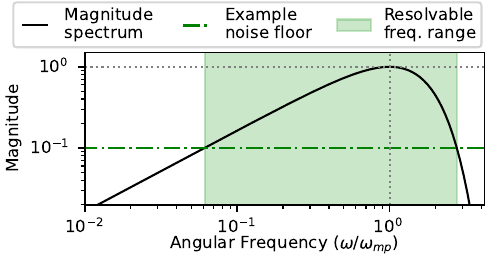}
			\caption{\label{fig:monopulse_spectrum} Magnitude spectrum of the Gaussian monopulse, normalized to its peak amplitude.  The monopulse's resolvable frequency range is roughly defined as the frequencies where this distribution is greater than the noise floor.   The highest SNR occurs at the peak  ($\omega / \omega_{mp} = 1$) and falls off on either side.  }
		\end{figure}

		The monopulse has two important parameters.  The first is the inverse of the width, $\omega_{mp} \equiv 1 / \sigma_{mp}$, which sets both the center frequency and frequency resolution of the pulse.  This concept is most easily understood by plotting the monopulse's magnitude spectrum,
		
		\begin{equation} \label{eq:monopuse_power_spectrum}
			\begin{split}
				\left| \mathcal{F} \left\{ g_{mp} \right\} \right| = a \frac{\omega}{\omega_{mp}} \, \text{exp} \left( \text{-}\frac{1}{2} \left(\frac{\omega}{\omega_{mp}}\right)^2 \right),
			\end{split}
		\end{equation}
				
		\noindent in Figure~\ref{fig:monopulse_spectrum}.  Here, $\mathcal{F} \left\{ \, \right\}$ is the Fourier transform.  This plot shows that Eq.~\ref{eq:monopuse_power_spectrum} is a distribution with a peak at $\omega=\omega_{mp}$ and a finite width.  SNR is roughly defined as the ratio of this distribution to a noise floor, and SNR $\gtrsim 1$ is roughly required to resolve the monopulse.  If the noise floor were, for example, 10\% of the spectrum's peak, then the resolvable frequency range would be approximately $0.06 < \omega / \omega_{mp} <2.8$ (shown in Figure~\ref{fig:monopulse_spectrum}) with the highest SNR at $\omega /\omega_{mp}=1$.  By default, we typically choose $\omega_{mp}$~$\scriptstyle \lessapprox$~$\omega_p$ as discussed further in Sec.~\ref{sec:uncertainty}. 
		
	 	The second important parameter is $\tau$, which is the time between sequential pulses.  The inverse of $\tau$ is the acquisition rate of the pulsed PIP system.  	 	
	 	The primary factor limiting the lower bound of $\tau$ is the time between the pulse and the settled reflected signal~\cite{dubois2021,brooks2023}. Without sufficient time between pulses, new outbound pulses would overlap with the reflection (or their ringdowns) from previous pulses.  A method of addressing this is to install the RFIV board in close proximity to the PIP in order to reduce the time delay between transmitted and reflected signals.

	\subsection{Pulsed calibration \label{subsec:pulsed_cal}}
	
		The calibration process for the swept and pulsed PIP approaches effectively use the same two calibration steps but with some additional nuance.  
		The first is that the 1st cal. plane is now at the oscilloscope (Figure~\ref{fig:swept_and_pulsed_setups}b).  
		The second is that the measurements of the 3 calibration standards are done differently: the ``truth'' measurements are made with a VNA but the 2nd calibration plane measurements are made with the pulsed hardware.  Interpolation is likely required to ensure that the frequency bases for the two measurements match.
		When measuring the RF calibration standards, SNR can be improved by averaging over multiple pulses.  
		As the precise value of $\omega_p$ is not typically known in advance, we often mitigate this unknown by calibrating and then operating with multiple values of $\omega_{mp}$.

	\subsection{Pulsed analysis \label{subsec:pulsed_analysis}}
	
		Because of the inherent tradeoff between time resolution and SNR when using the pulsed method, careful analysis of the measurements are particularly important.  There are several notable differences between the swept and pulsed methods.  
		
		The first is the careful selection of $\tau$ and $\omega_{mp}$ as discussed in Secs.~\ref{subsec:pulsed_setup} and~\ref{subsec:pulsed_cal}.  The second is using windowing functions.  Before applying Eq.~\ref{eq:IV_to_Z} to the raw measurements of $V_{RF}(t)$ and $I_{RF}(t)$, we typically apply a Hann window with a width on order of  $10^2 \, \sigma_{mp}$ to $10^3 \, \sigma_{mp}$ to each measured pulse to suppress the noise floor between the sequential pulses.  Optionally, sequential pulses can also be averaged. 
		
		Figure~\ref{fig:RFIV_pulses_realdata}b shows the calibrated $\Gamma_{pip}$ measurement and fit resulting from the raw data in Figure~\ref{fig:RFIV_pulses_realdata}a.
			
					
			
\section{Uncertainty analysis \label{sec:uncertainty}}

	In this work, we are arguing that our historical PIP analysis has not been optimized for SNR, and this has artificially lowered the PIP's upper-density limit, decreased measurement quality, and reduced acquisition rates.  In this section, we use a Monte Carlo (MC) uncertainty analysis~\cite{coleman1999} to quantify improvements in SNR due to our updated model (Sec.~\ref{sec:model}) and methods (Sec.~\ref{sec:methods}) and make recommendations for optimized PIP operation and analysis.

	\subsection{Methodology}

		In summary, our MC uncertainty analysis takes the following form. i) We develop a model for PIP measurements that includes Gaussian noise.  ii) We simulate a noisy PIP measurement for a fixed set of parameters and iii) perform a fit to calculate its plasma properties.  iv) We repeat steps ii and iii a statistically significant number of times (e.g. $10^4$), which allows the Gaussian distribution in the noise to propagate to each plasma parameter. v) We quantify the uncertainty in each plasma parameter by taking the standard deviation of each parameter across the multiple fits.  We then repeat this analysis for each set of desired hyperparameters (e.g. $\alpha'$, $\omega_{mp}$, etc).  Below we detail each step.
		
		In developing our model for a noisy PIP measurement, we start with the definition of the reflection coefficient (Eq.~\ref{eq:Gamma_definition}) which is the ratio of the is the reflected signal ($V_{refl}$) from the PIP to the transmitted ($V_{trans}$) signal to the PIP.  We assume that only $V_{refl}$ is susceptible to noise as $V_{trans}$ is measured at the VNA before being transmitted.  	
		Therefore, a noisy measurement of $\Gamma$ has the form,
		\begin{equation} \label{eq:Gamma_meas}
				\begin{split}
				\Gamma_{meas} (\omega)& = \frac{V_{refl}(\omega) + V_{noise}(\omega)}{V_{trans}(\omega)}\\
				& = \Gamma_{ideal}(\omega) + \Gamma_{noise}(\omega),
				\end{split}
		\end{equation}
		which includes the ideal PIP measurement ($\Gamma_{ideal}\equiv V_{refl} / V_{trans}$) plus noise ($\Gamma_{noise} \equiv V_{noise} / V_{trans}$).  
		For $\Gamma_{ideal}$, this analysis uses Eqs.~\ref{eq:Z_pip} or~\ref{eq:Z_diff}.  
		This analysis models $V_{noise}$, the numerator of $\Gamma_{noise}$, as a complex noise floor with the form,
		\begin{equation} \label{eq:V_noise}
			\begin{split}
				V_{noise}(\omega; \alpha) &= \mathcal{N}(0, \alpha) e^{ j U(0, 2 \pi)  }.
			\end{split}
		\end{equation}
		This expression's amplitude is a Gaussian or normal distribution, $\mathcal{N}(0, \alpha)$, with a zero mean and standard deviation, $\alpha$. 
		Eq.~\ref{eq:V_noise} has a uniformly random phase shift, $U(0,2\pi)$, between 0 and $2\pi$ with respect to $V_{trans}$. 
		For the swept method, we assume $V_{trans}(\omega)=a$, i.e. the transmitted signal has a constant amplitude, and therefore the noisy swept PIP model is
		\begin{equation} \label{eq:Gamma_noise_swept}
			\begin{split}
				\Gamma_{noise}(\omega; \alpha') = \mathcal{N}(0, \alpha') e^{ j U(0, 2 \pi)  },
			\end{split}
		\end{equation}
		where $\alpha' \equiv \alpha /a $ is our adjustable noise parameter and $1/\alpha'$ is effectively SNR.  
		For the pulsed method, we assume that $V_{trans}(\omega)$ is a Gaussian monopulse and therefore equal to Eq.~\ref{eq:monopuse_power_spectrum}.  The noisy pulsed PIP model is therefore
		\begin{equation} \label{eq:Gamma_noise_pulsed}
			\begin{split}
				\Gamma_{noise} (\omega; \alpha', \omega_{mp}) =  \frac{\mathcal{N}(0, \alpha') e^{ j U(0, 2 \pi)  }}{\frac{\omega}{\omega_{mp}} \text{exp} \left( -\frac{1}{2} \left( \frac{\omega}{\omega_{mp}} \right)^2 \right) }. 
		\end{split}
		\end{equation}
		Note that Eq.~\ref{eq:Gamma_noise_swept} is dependent on a single parameter, $\alpha'$, and Eq.~\ref{eq:Gamma_noise_pulsed} is dependent on two: $\alpha'$ and $\omega_{mp}$.
		
		With the noisy PIP models established, the next step is to simulate them.  
		For both swept and pulsed approaches, we define frequency, $\omega$, as a uniform set of $N$ discrete frequencies with step size, $\Delta \omega=(\omega_f - \omega_i)/(N-1)$,
		where the $\omega_i$ and $\omega_f$ are the initial and final frequencies of the sweep, respectively.  Unless otherwise specified, we use the following as default values for calculating $\Gamma_{meas}$ in the MC analysis: $\omega_p / 2\pi=100$ MHz, $\nu'=0.10$, $t_s' = 0.10$, $r_m = 0.5$ inch (12.7 mm), $\omega_f=\omega_p$, $\omega_i = \Delta \omega = \omega_f/N$, $N=10^3$,  $\alpha'=10^{-4}$, and $\omega_{mp} =  \omega_p$.  We choose these values because they are similar to our previous experimental measurements.
		
\begin{figure}
			\includegraphics[]{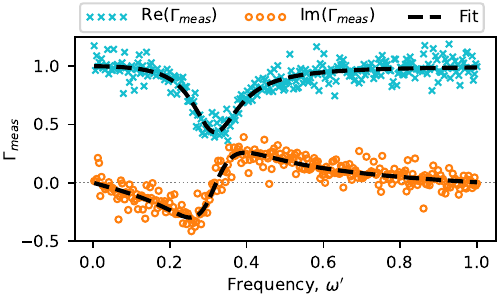}
			\caption{\label{fig:MC_example_single} Example of a single simulation of a noisy, swept PIP measurement ($\Gamma_{meas}$) with fit. $N=316$ and $\alpha'=10^{-1}$.} 
		\end{figure}

		Next, we fit our simulated $\Gamma_{meas}$ with our models ($\Gamma_{pip}$ or $\Gamma_{diff}$) to provide measurements of density ($n$), electron damping ($\nu$), and the sheath thickness ($t_{sh}$).  For fitting to $Z_{pip}$, we convert $\Gamma_{meas}$ to $Z$ with Eq.~\ref{eq:Z_to_Gamma}.  Figure~\ref{fig:MC_example_single}  shows an example simulated measurement and fit.  
		
		
		Next, we repeat the previous two steps $M=10^4$ times to generate $M$ fit values for $n$, $\nu$, and $t_{sh}$.  For each, we calculate the standard deviation ($\sigma$)  and divide by their ideal value to get the normalized uncertainty, $\sigma'$.  For density, this takes the form,
		\begin{equation} \label{eq:uncertainty_error}
			\begin{split}
				\sigma'_n =  \frac{\sigma_n}{n_{ideal}},
			\end{split}
		\end{equation}
		where $n_{ideal}$ is the ideal density value used to generate $\Gamma_{ideal}$.  In addition to uncertainty, we also calculate a normalized bias error, $\beta'$, for each plasma parameter.  For density, this would be	
		\begin{equation} \label{eq:bias_error}
			\begin{split}
				\beta'_n =  \frac{\left< n \right>}{n_{ideal}}
		\end{split}
		\end{equation}
		where $\left< n \right>$ is the average density from the fit results.  For our work, $\beta'$ is consistently multiple orders of magnitude smaller than $\sigma'$, and we therefore focus our analysis exclusively on $\sigma'$.  
		

	
	\subsection{Comparing analysis methods}
	
\begin{figure}
			\includegraphics[]{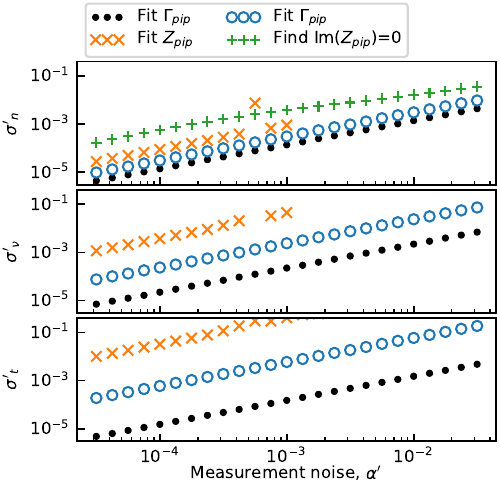}
			\caption{\label{fig:MC_analysis_method_comparison} Uncertainty results for four analysis methods, as a function of $\alpha'$.  Fitting with $\Gamma_{pip}$ results in the lowest uncertainty in all three plasma properties ($n$, $\nu'$, and $t'$) compared with the other three methods. }
		\end{figure}

		In this and our previous work~\cite{brooks2023}, we have alluded to several methods for extracting plasma properties from PIP measurements, which include: fitting with $\Gamma_{pip}$ (our recommended method), fitting with $Z_{pip}$, fitting with $\Gamma_{diff}$, and identifying the resonance at Im$(\Gamma_{diff})=0$ (for density only).  Our MC analysis allows us to compare the uncertainties associated with each method, and Figure~\ref{fig:MC_analysis_method_comparison} shows the results.  
		Note that for the Im$(\Gamma_{diff})=0$ method, we identified the zero intercept by iteratively applying a low-pass, forward-backward, Butterworth filter to our noisy $\Gamma_{diff}$ measurement with a decreasing corner frequency until a single zero intercept remained.  
		For this analysis, $\omega_f = 1.5 \omega_p$.
		
		Figure~\ref{fig:MC_analysis_method_comparison} shows that, of the four methods, fitting with $\Gamma_{pip}$ consistently provides the lowest uncertainty, by up to 2 orders-of-magnitude (for density and damping) and up to 4 orders-of-magnitude (for sheath thickness).  Note that the $Z_{pip}$ results are incomplete because low SNR caused fitting to become inconsistent above $\alpha'$ $\scriptstyle{ \gtrsim}$ $10^{-3}$. 
		
\begin{figure}
			\includegraphics[]{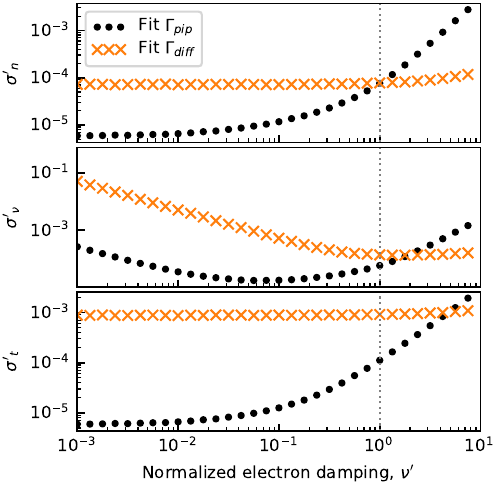}
			\caption{\label{fig:MC_analysis_total_vs_diff_with_nu} When damping is large ($\nu' \gtrsim 1$), fitting with $\Gamma_{diff}$ results in lower uncertainties than fitting with $\Gamma_{pip}$. }
		\end{figure}

		In the case of higher damping ($\nu'$ $\scriptstyle{ \gtrsim}$ $1.0$),  fitting with $\Gamma_{diff}$ results in lower uncertainty than $\Gamma_{pip}$ as shown in Figure~\ref{fig:MC_analysis_total_vs_diff_with_nu}.  This is because the magnitude of Im($Z_{diff}$) is less dependent on $\nu'$ than Im($Z_{pip}$) (see the discussion in Sec.~\ref{sec:model}). 
		
	\subsection{Improving SNR in $\mathbf{\Gamma_{pip}}$ analysis. }
	
		When fitting either swept and pulsed measurements with $\Gamma_{pip}$, there are a number of notable experimental and post-processing hyperparameter that impact measurement uncertainty: $N$, $\alpha'$, 
		$\omega_{mp}$, $\omega_i$, and $\omega_f$.  To investigate their impact, we perform the MC uncertainty analysis across a range of values for each hyperparameter.
		
\begin{figure}
			\includegraphics[]{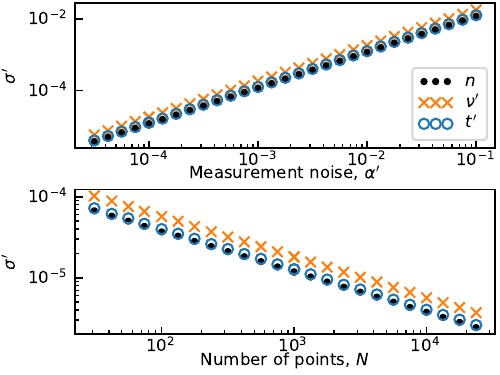}
			\caption{\label{fig:MC_analysis_a_vs_N_sigma} MC uncertainty analysis provides the uncertainty results of $n$, $\nu'$, and $t'$ for variable $\alpha'$, and $N$, 
			respectively, while holding the other values constant.  Power function fits show good agreement for most of the domains.}
		\end{figure}

		Figure~\ref{fig:MC_analysis_a_vs_N_sigma} shows the swept MC analysis for $N$ and $\alpha'$ 
		with several notable results.  First, the magnitude of the three uncertainties are consistently ordered:  $\sigma'_n$ $\scriptstyle{ \lesssim}$ $ \sigma'_{t}$  $\scriptstyle{ \lesssim}$ $\sigma'_{\nu}$.  Second, the uncertainties are linear in the log-log plot, meaning that each takes the form of a power function ($\sigma'$
		$\propto$
		$x^b$) where $b$ is a  scalar and $x$ is either $\alpha'$ or $N$. 
		We performed fits for all three plasma parameters and found that the exponent, $b$, is 1.00 and -0.50 
		for $\alpha'$ and $N'$, 
		respectively.  We also determined that these dependencies were independent of each other, and therefore the uncertainties for $\sigma_n'$, $\sigma_\nu'$, and $\sigma_t'$ are all proportional to
				
		\begin{equation} \label{eq:sigma_depedendency}
			\begin{split}
				\sigma' \propto \frac{\alpha'}{\sqrt{N}}.
			\end{split}
		\end{equation}
		
		\noindent The $\alpha'$ and $N$ dependencies are intuitive because uncertainty should decrease with lower $\alpha'$ (e.g. using more averaging) and with higher $N$ (adding additional measurement points).  
		Finally, we find the same relation in Eq.~\ref{eq:sigma_depedendency} for both swept and pulsed approaches.  
	
\begin{figure}
			\includegraphics[]{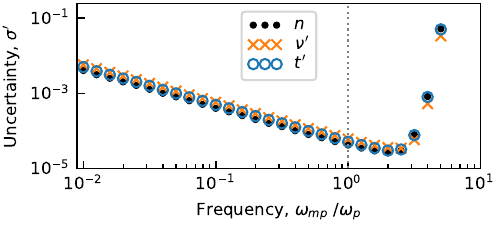}
			\caption{\label{fig:MC_sigma}  Pulse peak uncertainty analysis.  For the pulsed method, uncertainties have a minimum around $\omega_{mp}  \approx 2 \, \omega_p$ but also a steep rise around  $\omega_{mp}  \gtrsim 3 \, \omega_p$}
		\end{figure}

		The pulsed method introduces an additional parameter, $\omega_{mp}$, that sets the frequency resolution of the Gaussian monopulse.  Figure~\ref{fig:MC_sigma} shows the uncertainty results across a range of $\omega_{mp} / \omega_p$ and reveals that the uncertainties have a minimum at $\omega_{mp} \approx 2 \, \omega_p$ but also that there is a steep increase in uncertainty at $\omega_{mp} \gtrsim 3 \, \omega_p$.  As $\omega_p$ is generally not precisely known before a taking a measurement, we recommend choosing $\omega_{mp} \lesssim  \omega_p$.  These results are consistent across other values of $\nu'$ and $t'_{sh}$.  
		
\begin{figure}
			\includegraphics[]{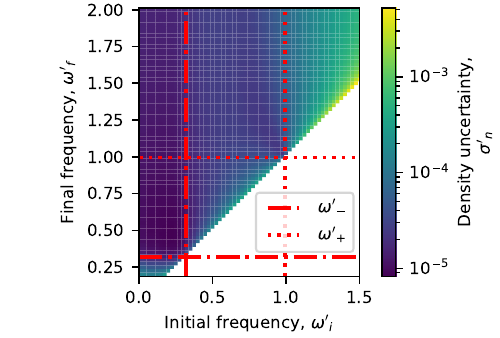}
			\caption{\label{fig:MC_analysis_fi_vs_ff} Uncertainty of $n$ over a range of $\omega_i$ and $\omega_f$ for a fixed $N$ and $\alpha$.  The lowest minima is the rectangular region ($ 0 < \omega'_i < \omega'_{-}$, $ \omega'_{-} < \omega'_f < \omega'_p$) and suggests that fitting only around $\omega'_{-}$ is required. }
		\end{figure}

		When using a VNA (i.e. the swept approach), we have independent control of $\omega_i$ and $\omega_f$.  Figure~\ref{fig:MC_analysis_fi_vs_ff} shows the uncertainty analysis of $\sigma'_n$ while varying both frequencies, and the two resonant frequencies, $(\omega'_{-}, \, \omega'_{+})$ = (0.318, 0.994), are indicated.  The results reveal two minima in $\sigma'_n$.

		The less prominent (higher) minimum occurs around ($\omega'_i$,~$\omega'_f$)~$\approx$~(1.0, 1.0) but for $\sigma_n$ only.  This is somewhat intuitive because $\omega'_+$ (Eq.~\ref{eq:plasma_resonance}) has a strong dependence on $\omega_p$, which relates to density (Eq.~\ref{eq:wp}).  
		
		The more prominent (lower) minimum is roughly bounded by the rectangular region ($ \omega'_i < \omega'_{-}$ and $ \omega'_{f} >  \omega'_-$) for $\sigma'_n$.  Analysis of $\sigma'_\nu$ and $\sigma'_t$ show this same minima.  The location and depth of this minima indicates that the best fit occurs when the measured sweep straddles $\omega_-$. This strongly suggests that $\omega_-$ sets the frequency resolution of the PIP rather than $\omega_p$ as previously believed.  We believe that this is likely due to the fact that $\omega_-$ is the location of Im($\Gamma_{pip}$)=0, the minimum of Re($\Gamma_{pip}$), and is surrounded by several inflection points and the largest magnitude of curvature (see Figure~\ref{fig:MC_example_single}).  Arguably, it is fitting to these features, in addition to normalizing with $\Gamma$, that leads to the most accurate results.  
		
		This finding has two interesting implications.  i) This suggests that the community has been spectrally over-resolving PIP measurements.  Resolving only up to $\omega_-$ (and not up to $\omega_p$) may effectively extend the PIP's upper density limit.  In this example, $\omega_-$ is roughly half an order-of-magnitude less than $\omega_p$, which relates to a full order-of-magnitude in density resolution (Eq.~\ref{eq:wp}).  Note that  $\omega_-$ is strongly dependent on $t'_{sh}$ (Eq.~\ref{eq:sheath_resonance}), and therefore this potential benefit only applies when the sheath is thin.  ii) Further reducing $\omega_-$ by deliberately decreasing $t'_{sh}$ (Eq.~\ref{eq:t_sh_prime}) has the potential to increase the PIP's density limit.  Possible methods to do this include: using a larger probe radius and electrically DC biasing the PIP at or near the plasma potential.

	\subsection{Uncertainty analysis summary}
	
		In summary, we recommend the following best practices to maximize PIP SNR.  We recommend fitting $\Gamma_{pip}$ to measurements if $\nu' < 1$ but fitting with $\Gamma_{diff}$ if $\nu' > 1$.  For $N$ with the swept method, we recommend using a reasonably large number (e.g. $10^3$ or $10^4$) that still provides an adequate acquisition rate.  While we do not have direct control over noise, $\alpha'$, we can use techniques such as averaging to reduce it.  
		For $\omega_{mp}$, we nominally recommend $\omega_{mp} = \omega_{p}$, but also recommend using multiple values to safeguard against an uncertain $\omega_{p}$.  
		
		An interesting result from this section is that $\omega_-$ appears to be the frequency that sets the upper density limit of the PIP, and not $\omega_p$ as previously believed.  This effectively extends the PIP's density limit, particularly when the sheath is ``thin''.  Therefore when choosing $\omega_i$ and $\omega_f$, we recommend values that straddle $\omega_-$ while also adjusting these values appropriately to safeguard against variability in $\omega_-$.

\section{Conclusions}

	The goal of this work has been to provide an approach to PIP analysis that increases measurement SNR in order to improve overall measurement quality, extend the upper density limit, increase acquisition rates, and allow the probe to be more accessible in the laboratory environment.  To achieve this, we presented our PIP design and model, our approach to PIP operation, calibration, and analysis, and finally a Monte Carlo uncertainty analysis that identified a number of operational and analysis steps that collectively improve SNR by multiple orders of magnitude.  We additionally provided evidence that the sheath resonance (and not the plasma frequency as previously believed) effectively sets the PIP's upper density limit, which has implications in additionally extending the PIP's upper-density limit.  
	
	In this work, we have also identified several paths for future work.  These include:  identifying the new value of the PIP's upper density limit and actively reducing the normalized sheath thickness to further increase the upper-density limit.  Finally, we believe that PIPs are an ideal candidate for educational plasma laboratories, particularly with the availability of commercial VNAs on the internet that are under \$100 and capable of resolving frequencies $>1$ GHz.

\section*{Acknowledgments}

	This work was supported by NRL base funding.  
	JWB wishes to thank his mentors, Mike McDonald and Erik Tejero, for their assistance, without which this work would not be possible.
	JWB wishes to thank Jeffrey Bynum for his assistance with technical drawings.
	JWB also wishes to thank his late teacher, Dr. Hugh W. Coleman~\cite{coleman1999}, for teaching him the importance of quantifying measurement uncertainty.  

\section*{Data Availability}

	The data that supports the findings of this study are available on Zenodo~\cite{brooks_code_nodate}.

\appendix


\section{The two-port network model for the PIP's stem \label{app:2port_cal}}

	In order to calibrate the PIP's stem, we construct a two-port model of the stem using an ideal, lossless transmission line (TL) model~\cite{caspers2011,pozar2011}, which are expressed in S (scattering) parameters in the following form,
	
	\begin{equation}
	\label{eq:ideal_TL_model_S}
	\begin{bmatrix}
	S_{11}(\omega) & S_{12}(\omega) \\ 
	S_{21}(\omega) & S_{22}(\omega)
	\end{bmatrix}
	=
	\begin{bmatrix}
	0 & e^{- j \omega L / v_p} \\ 
	e^{- j \omega L / v_p} & 0
	\end{bmatrix} .
	\end{equation}
	\noindent Here $L$ is the length of the TL,  $j$ is the imaginary number, $\omega$ is the angular frequency, $v_p= c / \sqrt{\epsilon_r}$ is the velocity of propagation of the transmission line, $\epsilon_r$ is the published relative dielectric constant of the TL, and $c$ is the speed of light in vacuum.  The resulting matrix can then be passed to \emph{scikit-rf}'s \emph{Network} function~\cite{scikitrf2023}.  Note that Eq.~\ref{eq:ideal_TL_model_S} is only accurate when $L$ is much smaller than the attenuation scale-length of the TL's dielectric.  If this condition is violated, an attenuation (i.e. loss) term could be inserted into Eq.~\ref{eq:ideal_TL_model_S}.  
				
\section{Example calibration \label{app:calibration_example}}

\begin{figure}[hbt!]
		\includegraphics[]{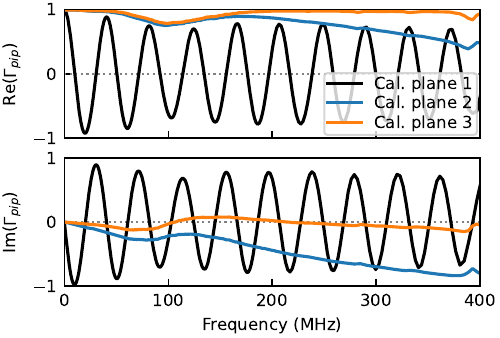}
		\caption{\label{fig:calibration_example} Example measured signals at each calibration plane (cal. plane):  the uncalibrated PIP measurement (cal. plane 1), partially calibrated measurement (cal. plane 2) and fully calibrated measurement (cal. plane 3).  }
	\end{figure}

	Figure~\ref{fig:calibration_example} shows an uncalibrated PIP measurement and how the measurement visually changes as each calibration step is applied.  Cal. plane~1  is the raw, uncalibrated PIP measurement.  Its ``oscillatory'' appearance is due to the phase shift caused by the length of the transmission line between the VNA and the PIP-monopole.  Cal. plane~2 is the partially calibrated PIP measurement that still contains contributions from the PIP and its stem and shows a small phase shift ($<2\pi$).  Cal. plane~3 shows the fully calibrated measurement, where both $\omega_{\pm}$ are apparent.  The calibration process is detailed in Sec.~\ref{subsec:calibration}.
	
\nocite{*}
\bibliography{biblio}

\end{document}